\documentclass[useAMS,usegraphicx,usenatbib]{mn2e}

\usepackage{amssymb}
\usepackage{times}
\usepackage{epsfig}

\usepackage{longtable}

\newcommand{\be}{\begin{equation}}
\newcommand{\ee}{\end{equation}}
\newcommand{\beq}{\begin{eqnarray}}
\newcommand{\eeq}{\end{eqnarray}}
\newcommand{\beqno}{\begin{eqnarray*}}
\newcommand{\eeqno}{\end{eqnarray*}}
\newcommand{\bitmz}{\begin{itemize}}
\newcommand{\eitmz}{\end{itemize}}

\def\msun{{\rm M}_{\odot}}

\def\Mdot{\dot{M}}

\def\mdot{\dot{m}}

\def\Mdotedd{\dot{M}_{\rm Edd}}
\def\Ledd{L_{\rm Edd}}

\def\Lsoft{L_{\rm soft}}
\def\Lx{L_{\rm X}}
\def\Labs{L_{\rm abs}}

\def\Tin{T_{\rm in}}

\def\NH{N_{\rm H}}

\def\epiv{E_{\rm p}}

\def\chired{\chi^2_{\rm red}}
\def\ergs{{\rm erg\,s^{-1}}}

\def\araa{ARA\&A}%
\def\apj{ApJ}%
\def\apjl{ApJ}%
\def\aap{A\&A}%
\def\mnras{MNRAS}%
\def\pasj{PASJ}%
\def\apjs{ApJS}%

\title[Spectral variability of ultraluminous X-ray sources]{Spectral variability of ultraluminous X-ray sources}
\author[Jari J. E. Kajava and Juri Poutanen]{Jari J. E. Kajava\thanks{E-mail:
jari.kajava@oulu.fi (JJEK), juri.poutanen@oulu.fi (JP)} and Juri Poutanen\footnotemark[1] \\
Department of Physical Sciences, Astronomy division, P.O.Box 3000, 90014 University of Oulu, Finland}
\begin{document}

\date{Accepted 2009 June 07. Received 2009 June 07; in original form 2008 November 21}

\pagerange{\pageref{firstpage}--\pageref{lastpage}} \pubyear{2009}

\maketitle

\label{firstpage}

\begin{abstract}
We study spectral variability of 11 ultraluminous X-ray sources (ULX) using archived {\it XMM-Newton} and {\it Chandra} observations. We use three models to describe the observed spectra: a power-law, a multi-colour disc (MCD) and a combination of these two models. We find that 7 ULXs show a correlation between the luminosity $\Lx$ and the photon index $\Gamma$. Furthermore, 4 out of these 7 ULXs also show spectral pivoting in the observed energy band. We also find that two ULXs show an $\Lx-\Gamma$ anti-correlation. The spectra of 4 ULXs in the sample can be adequately fitted with a MCD model. We compare these sources to known black hole binaries (BHB) and find that they follow similar paths in their luminosity-temperature diagrams. Finally we show that the `soft excess' reported for many of these ULXs at $\sim 0.2$ keV seems to roughly follow a trend $\Lsoft \propto T^{-3.5}$ when modelled with a power-law plus a `cool' MCD model. This is contrary to the $L \propto T^{4}$ relation that is expected from theory and what is seen for many accreting BHBs. The observed trend could instead arise from disc emission beamed by an outflowing wind around a $\sim 10$ solar mass black hole.
\end{abstract}


\section{Introduction}

Ultraluminous X-ray sources (ULX) are point like, non-nuclear (not an active galactic nucleus, AGN) objects with X-ray luminosities exceeding the Eddington limit for stellar mass black holes (StMBH), $\Ledd \approx 1.3 \times 10^{39} (M/10\msun)\, \ergs$. Many ULXs have been identified as accreting black holes because of their short- and long-term variability \citep[][and references therein]{MC04}. More than 300 ULXs have been discovered in nearby galaxies \citep{L08} and all types of galaxies have ULXs \citep{M06}. Although they were discovered with the {\it Einstein} mission almost thirty years ago \citep[see][ for review]{F89}, the nature of these sources is not yet well understood. 

Assuming that ULXs are isotropic emitters, the most luminous source M82 X-1 has a luminosity of $\Lx \sim 10^{41} \, \ergs$ \citep{PZ06}. Therefore, for such a high luminosity ULX the accreting black hole should be an intermediate-mass black hole (IMBH) with $M \gtrsim 1000 \msun$ not to exceed the Eddington limit. One argument used to support this idea is the presence of a $\sim 0.2$ keV `soft excess' in the spectra of several luminous ULXs \citep{KCPZ03,MF03}, which could be interpreted as a signature of a `cool' accretion disc around an IMBH. It is also possible, however, that this excess arises from the emission at a large distance from the central source \citep{K01,PLF07} or, maybe, is an artifact of a complicated absorber or emission from some thermal plasma surrounding the source \citep{GS06}. Another argument used in favour of the IMBH interpretation is the detection of low-frequency quasi-periodic oscillations (QPO) from ULXs in M82, Holmberg IX and NGC 5408 \citep{SM03,DGR06,SMW07}. However, similar QPO frequencies have been also detected from Cyg X-1 \citep{VCG94a}, which is certainly not an IMBH. 

There are also several arguments against the hypothesis that ULXs are powered by accretion onto IMBHs. The main issue is the formation of such massive objects in `normal' stellar evolution. Black holes of masses greater than $80 \msun$ are not expected to form even in the lowest metallicity environments and at solar metallicities the upper limit is about $15 \msun$ \citep{BBF09}. Furthermore, about 30 per cent of ULXs are associated with star formation regions \citep{M06} and a small number of them have been associated with young stellar counterparts \citep{M06,RLG08}. These facts can be naturally explained if ULXs are powered by accretion onto StMBHs.

\begin{table*}
 \centering
 \begin{minipage}{150mm}
  \caption{Observation log}
  \begin{tabular}{@{}llllll@{}}
  \hline
   ULX name  & Alternate & Coordinate & Distance\footnote{Distance estimates are from \citet{LB05} except for HoII ULX-1 and NGC 5408 ULX-1, where the distances are taken from \citet{KDG02} and \citet{LM05}, respectively.} & \multicolumn{2}{c}{Observation IDs} \\
   			 &	names	 & J2000 	  & Mpc		 & {\it XMM-Newton} & {\it Chandra}	\\
 \hline
 NGC 253 X-2 	& NGC 253 PSX-2	& 00:47:33.0 $-$25:17:49 & 3.0 	& 0110900101 0125960101 0125960201 	& 790\\
			 	& NGC 253 ULX2 	& 			 			 &		& 0152020101 0304850901 0304851001  \\
			 	& NGC253 XMM1	& 			 			 &		& 0304851201 0304851101 \\
 NGC 253 X-4	& NGC 253 PSX-7	& 00:47:35.3 $-$25:15:12 & 3.0 	& 0110900101 0125960101 0152020101  & 790 3931 \\
			 	& NGC253 XMM3	& 			 			 & 	 	& 0304850901 0304851001				&   \\

 NGC 253 X-9 	& NGC 253 PSX-5	& 00:47:22.6 $-$25:20:51 & 3.0 	& 0110900101 0125960101 0152020101 	& 3931  \\
			 	& NGC253 XMM2	& 			 			 &		& 0304850901 0304851001 0304851201  \\

 NGC 1313 ULX-1 & NGC1313 XMM1	& 03:18:19.8 $-$66:29:10 & 3.7 	& 0106860101 0150280301 0150280401  &\\
			 	& IXO 7			& 			 			 &	 	& 0150280601 0150281101 0205230201  &\\
			 	&				& 			 			 &	 	& 0205230301 0205230401 0205230501  &\\
			 	&				& 			 			 &	 	& 0205230601 0405090101				&\\

 NGC 1313 ULX-2 & NGC1313 XMM3	& 03:18:22.0 $-$66:36:04 & 3.7 	& 0106860101 0150280101 0150280301   &\\
			 	& IXO 8			&  			 			 &	 	& 0150280401 0150280501 0150280601   &\\
			 	&				&  			 			 &	 	& 0150281101 0205230201 0205230301   &\\
			 	&				&  			 			 &	 	& 0205230401 0205230501 0205230601   &\\
			 	&				&  			 			 &	 	& 0301860101 0405090101 			& \\

 IC 342 X-6 	& IC 342 ULX1	& 03:45:55.7 $+$68:04:55 & 3.9 	& 0093640901 0206890101 0206890201  & 2916 2917 \\
			 	& IXO 22		&  			 			 &	 	& 0206890401 						& \\

 HoII ULX-1		& IXO 31		& 08:19:29.0 $+$70:42:19 & 3.39	& 0112520601 0112520701 0112520901  & \\
			 	&				&  			 			 &	 	& 0200470101 & \\

 HoIX ULX-1 	& NGC 3031 ULX2	& 09:57:54.1 $+$69:03:47 & 3.42 & 0112521001 0112521101 0200980101 & 3786 3787 3788  \\
			 	& IXO 34		&  			 			 &	 	&  									& \\

 NGC 5204 ULX-1 & NGC5204 XMM1	& 13:29:38.6 $+$58:25:06 & 4.3 	& 0142770101 0142770301 0150650301 & 3933   \\
			 	& IXO 77		&  			 			 &	 	& 0405690101 0405690201 0405690501 & \\

 NGC 5408 ULX-1 & NGC5408 XMM1	& 14:03:19.6 $-$41:23:00 & 4.8  & 0112290501 0112290601 0112290701 	&   \\
			 	&				&  			 			 &	 	& 0112291201 0302900101  			&\\

 NGC 6946 X-6 & NGC 6946 ULX3	& 20:35:00.7 $+$60:11:31 & 5.5 	& 0200670301 0401360301 			& 1043 4404 4631  \\
			 	&				&  			 			 &	 	&    								& 4632 4633\\

\hline
\end{tabular}
\end{minipage}
\end{table*}

There are numerous models put forward on how the luminosity of a StMBH can exceed its Eddington limit. Truly super-Eddington models include the advective slim disc models \citep{JAP80,A88,Kaw03}, which have been successfully used to describe some ULX spectra \citep{VM06}. Accretion discs with strong outflows \citep{SS73,K01,KP02,PLF07} can also explain the high luminosities of ULXs. Other plausible scenarios involving StMBHs that can produce the high observed luminosities include the photon bubble model \citep{A92,B01} and beaming of radiation by a relativistic jet \citep{KFM02,FKSB06}. Strong beaming, however, can be ruled out in some cases by observations of photo-ionised nebulae surrounding the ULXs because the observed line properties require luminosities comparable to the observed X-ray luminosities \citep{PM03,KWZ04,AFSA07,KC09}.

Because of the limited observing band of the current X-ray observatories capable of detecting ULXs ($\sim$ 0.2$-$10 keV), it is not surprising that many of the models mentioned above have been successfully used to model the observed spectra of some particular ULX. This is mainly because many previous studies have been concentrated only on modelling single observations of individual ULXs \citep[e.g.][]{FK05,WMR06,SRW06}. The lack of multi epoch observations can lead to misinterpretation of the physical meaning of the derived spectral parameters. By studying the spectral variability of ULXs, we hope to learn more about the physical nature of the specific spectral components by comparing their variations to expectations from theoretical models. Some previous works have studied spectral variability of ULXs \citep{FZ03,DM04,FeKa06b,FeKa06,RKW06,BWCR08}, but the samples of ULX with multiple observations have been small.\footnote{Note a recent paper by \citet{FK09}, who consider a large sample of ULXs.} Interestingly, the study by \citet{FeKa06} showed that the two ULXs in NGC 1313 galaxy have different types of spectral variability. Their subsequent study \citep{FK07} also showed that the soft excess in NGC 1313 ULX-2 did not follow the $L \propto T^{4}$ relation that is expected from theory when modelled with a multi-colour disc (MCD). Therefore, in order to study the spectral variability of ULXs in better detail, we have selected a sample of 11 ULXs that have been observed multiple times with {\it XMM-Newton} and {\it Chandra}. In this paper, we present the results of spectral analysis of these data sets.

\section{Observations and data reduction}

We searched the data archives for high S$/$N {\it XMM-Newton} and {\it Chandra} observations of ULXs that have been observed multiple times. We also required that the observations were not effected by photon pile-up. There were 11 sources that met the requirements: NGC 253 X-2, NGC 253 X-4, NGC 253 X-9, IC 342 X-6, NGC 1313 ULX-1, NGC 1313 ULX-2, Holmberg II ULX-1, Holmberg IX ULX-1, NGC 5204 ULX-1, NGC 5408 ULX-1 and NGC 6946 X-6 (see Table 1). In the cases where sources have multiple names, we followed the naming convention of \citet{LB05}. 

{\it XMM-Newton} data reduction was done using the SAS version 7.1.0. In order to produce a homogeneously calibrated data set, we processed all the observations using the observation data files with the latest calibration files as of March 2008. The reduction was done according to the User's guide to the {\it XMM-Newton} Science Analysis System.\footnote{Guide is available at \\ http://xmm.esac.esa.int/external/xmm\_user\_support\\/documentation/sas\_usg/USG.pdf.} Specifically, we used the event selections "FLAG$==$0 \&\& PATTERN$<=$4" and "\#XMMEA\_EM \&\& PATTERN$<=$12" for the EPIC-pn and EPIC-mos instruments, respectively. {\it Chandra} observations were reduced using the CIAO 4.0.1 software. We processed all the observations with calibration files from CALDB 3.4.3 following the {\it Chandra} science threads.\footnote{Threads are available at http://cxc.harvard.edu/ciao\\/threads/index.html.} 

We used circular source extraction regions of varying radii (between 20'' to 40'' for {\it XMM-Newton} and 4'' to 12'' for {\it Chandra}). Close-by source free regions were used as a background.\footnote{In the case of {\it XMM-Newton}, the background region was chosen from the same EPIC-pn CCD chip as the source.} The resulting background subtracted spectral data were binned to require at least 20 counts per bin to ensure adequate statistics for the XSPEC spectral fitting. To ensure the best possible spectral quality, the time periods of high background were eliminated from the data for both {\it XMM-Newton} and {\it Chandra} observations. In the case of {\it XMM-Newton} the exclusion was done by removing the time periods where the 10$-$12 keV full field count rate was above 0.4 cts s$^{-1}$ and 0.35 cts s$^{-1}$ for EPIC pn and mos, respectively. For {\it Chandra}, we used the {\sc analyze\_ltcrv.sl} script on a source free event file to remove the high background periods.

\begin{figure}
\centerline{\epsfig{file= 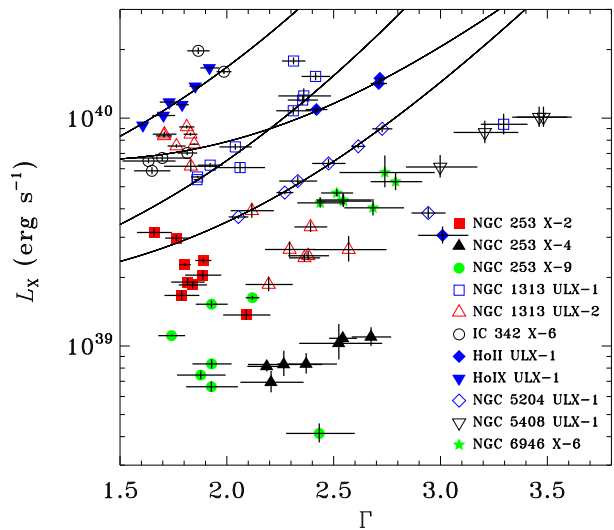,width=8.0cm}}
\caption{The intrinsic 0.4$-$10 keV band luminosity versus the photon index obtained from the {\sc powerlaw} model fits. Different observations of a given object showing spectral pivoting in the observed band are connected by a theoretical curve (\ref{Fgamma}). Some sources show correlation, but no spectral pivoting and some other even show anti-correlation. NGC 1313 ULX-1, Holmberg II ULX-1, Holmberg IX ULX-1, NGC 5204 ULX-1 show $\Lx-\Gamma$ correlation {\em with} spectral pivoting in the observed band. 
NGC 253 X-4, IC 342 X-6 and NGC 5408 ULX-1 show $\Lx-\Gamma$ correlation {\em without} spectral pivoting, and 
NGC 253 X-2 and NGC 1313 ULX-2 are $\Lx-\Gamma$ anti-correlated sources. 
NGC 253 X-9 shows no $\Lx-\Gamma$ correlation and NGC 6946 X-6 is not variable enough to make firm conclusions.}
\label{pofitresults}
\end{figure}

\section{Spectral analysis and results}

We used three models to describe the observed spectra: a {\sc powerlaw}, a MCD ({\sc diskbb} model in XSPEC) and a combination of these two models. The effect of the interstellar medium was taken into account with the {\sc wabs} model in XSPEC. The hydrogen column density was let as a free parameter, but was required to have at least the Galactic value \citep{DL90}. In some cases, none of these simple models gave statistically acceptable fits to the data. In most of these cases, the fits could have been improved by adding more model components like gaussians at $\lesssim 1$ keV (NGC 5204 ULX-1, OBSID 3933, see \citet{RKW06}; NGC 5408 ULX-1, OBSID 0302900101) or by changing metal abundances of the absorber (HoII ULX-1, see \citet{GRR06}). We checked that our best fitting parameters did not change significantly by adding these components, so we decided to ignore these facts, because we wanted to keep our models as simple as possible. 

We fit the spectra in the 0.4$-$10 keV band (or up to energies which are not noise dominated). XMM-Newton provides spectral coverage down to $\sim 0.2$ keV, but we decided to restrict our fitting range to 0.4 keV because of the known calibration uncertainties below this value.\footnote{See EPIC status of calibration and data analysis XMM-SOC-CAL-TN-0018.} All the parameter errors in the following figures are given at the $90$ per cent confidence level. The errors for the fluxes were computed using the {\sc cflux} convolution model in XSPEC. In the case of the {\sc powerlaw} model, the luminosities were calculated for the 0.4$-$10 keV band. The MCD component luminosities are bolometric.

\subsection{Power-law type ULXs and the $\Lx-\Gamma$ correlation} \label{po_type}

A simple absorbed {\sc powerlaw} model provides a good fit for most of the observations in our sample (see Table \ref{bestfits}). We find a correlation between the luminosity and the photon index (hereafter $\Lx-\Gamma$ correlation) for most of the ULXs, see Fig. \ref{pofitresults}. Similar correlations were also observed for NGC 1313 ULX-1 \citep{FeKa06} and Antennae X-11 \citep{FeKa06b}. The ULXs with the $\Lx-\Gamma$ correlation are NGC 253 X-4, IC 342 X-6, NGC 1313 ULX-1, Holmberg II ULX-1, Holmberg IX ULX-1, NGC 5204 ULX-1 and NGC 5408 ULX-1. In contrast, two ULXs of the sample, NGC 253 X-2 and NGC 1313 ULX-2, show an anti-correlation between the luminosity and $\Gamma$ \citep[see Fig. \ref{pofitresults} and ][for comparison]{FeKa06b,FeKa06}. We also find that the luminosity of NGC 253 X-9 is clearly not correlated with $\Gamma$ and that NGC 6946 X-6 is not significantly variable to make firm conclusions. 

For NGC 1313 ULX-1, Holmberg IX ULX-1 and NGC 5204 ULX-1, the modelled absorption column $\NH$ is also correlated with $\Gamma$ \citep[see also fig. 1 in][for NGC 5204 ULX-1]{FK09}. If this correlation has no physical origin, then the luminosities of the softest spectra would be `artificially boosted'. To evaluate this effect on the derived {\it intrinsic} luminosities, we fixed the absorption columns to their mean values, and fitted the data again. This lead to somewhat worse fits for the cases where the best fitting absorption columns differ most from the mean, but the $\Lx-\Gamma$ correlations remained. Also, the observed (i.e. absorbed) luminosities $\Labs$ are correlated with $\Gamma$, but the $\NH - \Gamma$ dependence leads to a `shallower' correlation between $\Labs$ and $\Gamma$. Therefore, we conclude that the $\Lx-\Gamma$ correlations are real.  

If the $\NH - \Gamma$ correlation is indeed artificial, then there must be a low-energy break in the power-law component below $\sim 1$ keV in these spectra. This could be then be interpreted as a sign that this component arises from Comptonization of the disc photons in a `corona' surrounding the source. It is also possible, however, that the correlation between $\NH$ and $\Gamma$ has a physical origin, for example a growing luminosity might lead to a stronger outflow and therefore to a denser environment.

\begin{figure}
\centerline{\epsfig{file= 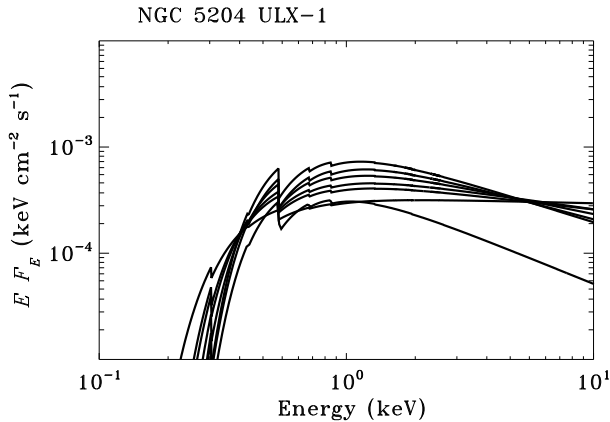,width=8.0cm}}
\caption{All observed $EF_E$ spectra of NGC 5204 ULX-1 fitted with an absorbed {\sc powerlaw} model. The curves show the deconvolved model spectra. Spectral pivoting occurs at $\sim 5.5$ keV.}
\label{pivoting}
\end{figure}

In general, all the sources that have luminosities {\em permanently above} $\sim 3 \times 10^{39} \, \ergs$ show a power-law type spectra. In contrast, the three sources with luminosities {\em permanently below} $\sim 3 \times 10^{39} \, \ergs$ have non-power-law (or thermal) type spectra (see next subsection). An exception to this `rule' is NGC 1313 ULX-2 which has non-power-law type spectra at high luminosities and power-law type spectra at lower luminosities very much like the very high state spectra of XTE J1550-564 \citep{KD04}. In some sources, both {\sc powerlaw} and {\sc diskbb} models give statistically acceptable (or equally good) fits. We assign them to a non-power-law type because their variability is consistent with Galactic black hole binaries (BHB) when modelled using {\sc diskbb}. Also, a simple absorbed {\sc powerlaw} model does not provide statistically acceptable fits for NGC 6946 X-6, but the {\sc diskbb $+$ powerlaw} model gives good fits for most observations (see Sect. \ref{soft_excess} and Table \ref{bestfits}). We therefore assign this source to a power-law type.

Four out of the 7 ULXs with the $\Lx-\Gamma$ correlation (NGC 1313 ULX-1, Holmberg II ULX-1, Holmberg IX ULX-1 and NGC 5204 ULX-1) show also signs of spectral pivoting. Similar pivoting has been also observed in Cyg X-1 \citep{Z02} and several AGNs \citep[see e.g.][ table 1, and references therein]{Z03}. 
 The pivoting is clearly seen for NGC 5204 ULX-1 at $\sim 5.5$ keV (see Fig. \ref{pivoting}) and the three high luminosity observations of Holmberg II ULX-1 seem to pivot at $\sim 3.5$ keV. The pivoting in NGC 1313 ULX-1 and Holmberg IX ULX-1 occur at the boundary of the {\it XMM-Newton} energy range at $\sim 9$ and $10$ keV, respectively. To evaluate whether these pivots are real, we also show in Fig. \ref{pofitresults} the dependences between the luminosity in a given energy band ($E_1,E_2$) and the photon index expected for a pivoting power-law \citep{Z03}:
\be \label{Fgamma}
L_{E_1 - E_2} = 4\pi D^2 F_{E_1 - E_2} = C \epiv^{\Gamma} \frac{E_2^{2 - \Gamma} - E_1^{2 - \Gamma}}{2-\Gamma},
\ee
where $D$ is the distance, $F_{E_1 - E_2}$ is the flux in the selected energy band and $C$ is a constant. Equation (\ref{Fgamma}) gives a good representation of the observed $\Lx-\Gamma$ relation for all these pivoting ULXs. We therefore conclude, that the pivoting is most likely real even in the cases of NGC 1313 ULX-1 and Holmberg IX ULX-1. However, when the spectrum gets very soft at $\Gamma \gtrsim 3$, NGC 1313 ULX-1, Holmberg II ULX-1 and NGC 5204 ULX-1 also show deviations from this dependence. These deviations to the low/soft states \citep{DM04}, are similar to those seen in Galactic BHBs \cite[see][ fig. 16a]{Z02}.

The process producing the observed power-law type spectrum is most likely Compton up-scattering of soft disc photons in a hot plasma surrounding the inner part of the disc. The observed $\Lx-\Gamma$ correlation and the pivoting in the sources can be explained by variability of the luminosity of the soft disc photons. A larger disc luminosity leads to a softer X-ray spectrum \citep{Z03}. 
An example of such variability for a constant hot-plasma luminosity is given in fig. 3 of \citet{ZG01} for 3C 120. The similarity of this source to NGC 5204 ULX-1 (and the other 3 ULXs with pivoting) is remarkable (see Fig. \ref{pivoting}). 
Transition to the low/soft state can be associated with the decreasing luminosity of the hot plasma (corona) with the nearly constant disc luminosity.

The highest quality data also seems to show a break/cutoff in the power-law component above $\sim 3$ keV \citep{SRW06,GRD09}. This could be then interpreted as a sign of a cool and optically thick corona \citep{GRD09}. Unfortunately the limited observing band of {\it XMM-Newton} and {\it Chandra} and the typical data quality in our sample do not allow us to see these high energy breaks/cutoffs in most of these data. A dedicated monitoring campaing of these power-law type ULXs would help us to check whether these breaks are always present, which would be highly useful in order to understand the nature of the emission mechanisms in ULXs.

\subsection{Non-power-law type ULXs}

\begin{figure}
\centerline{\epsfig{file= 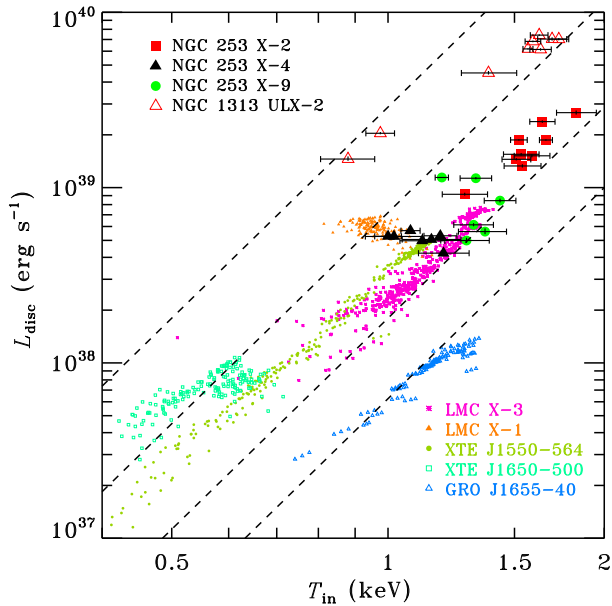,width=8.0cm}}
\caption{The MCD model luminosity-temperature diagram for the non-power-law type ULXs. All the sources roughly follow the tracks of known BHBs shown by small symbols.}
\label{dbbfits}
\end{figure}

Out of the  selected 11 ULXs, four ULXs can be adequately fitted with an absorbed MCD model. These sources include the two ULXs that show the $\Lx-\Gamma$ anti-correlation (NGC 253 X-2 and NGC 1313 X-2) and the two low-luminosity ULXs (NGC 253 X-4 and NGC 253 X-9). The resulting fits are shown in a luminosity-temperature (hereafter $L-T$) diagram in Fig. \ref{dbbfits} together with the observations on known BHBs adopted from \citet{GD04}. These four ULXs follow the behaviour seen in BHBs. The properties of the sources can be summarised as follows.

\begin{itemize} 
\item[(i)] NGC 253 X-2 closely follows the $L \propto T^{4}$ relation. Its track on the diagram is a high luminosity--high temperature extension to XTE J1550-564 and LMC X-3. The relatively high luminosities and hot inner disc temperatures of this source compared to BHBs could be explained by assuming that this source is a maximally rotating Kerr black hole with $M \sim 70 \msun$ and inclination of $\sim 70$ degrees \citep{HK08}.

\item[(ii)] NGC 253 X-4 on the other hand behaves like LMC X-1 which clearly does not follow the expected $L \propto T^{4}$ relation. As discussed in \citet{PLF07}, this can be understood if these sources are StMBHs (or even neutron stars) accreting slightly above their Eddington accretion rates. In this case the observed temperature corresponds to the emission from the spherization radius or even the distant photosphere if observed edge-on.

\item[(iii)] NGC 253 X-9 first seems to follow the $L \propto T^{4}$ track at lower luminosities, but then at $L \gtrsim 10^{39} \, \ergs$ deviates towards lower temperatures similarly to LMC X-1 and NGC 253 X-4. This type of behaviour is also seen in the spectra of XTE J1550-564 when it is in the very high state (VHS) instead of the high/soft state \citep{KD04}.

\item[(iv)] NGC 1313 X-2 shows a slight deviation from the $L \propto T^{4}$ track much like XTE J1650-500. As the luminosity of this source greatly exceeds the Eddington limit for a 10 solar mass black hole, the simple {\sc diskbb} model is not anymore a good approximation of the spectrum. Indeed the {\sc diskbb} model failed to properly model the high-energy curvature of the spectrum, predicting slightly too steep cutoff. Because at lower luminosities the observed spectra of this source is very much like VHS spectra of XTE J1550-564 \citep{KD04}, the higher luminosity non-power-law type spectra could be understood as an ultraluminous state \citep{R07,S07}. However, a more elaborate model than the {\sc diskbb} should be used to investigate this state further. 
\end{itemize}

\begin{figure}
\centerline{\epsfig{file= 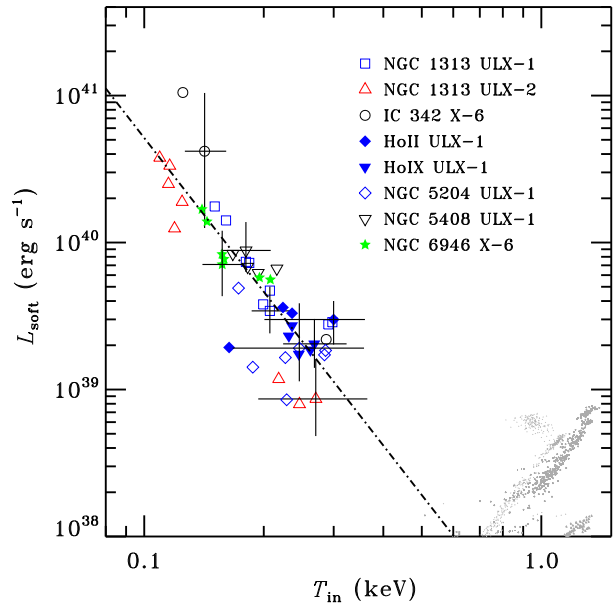,width=8.0cm}}
\caption{The luminosity-temperature diagram of the soft excess from the {\sc powerlaw} plus cool MCD model fits (only one set of errors per source is shown for clarity). All the sources with this model follow roughly a $\Lsoft \propto \Tin^{-3.5}$ scaling. Note however that this scaling is subject to some modelling uncertainties (see subsection \ref{soft_excess}) and that the coolest and highest luminosity soft excesses of NGC 253 X-2 are artifacts of an improper modelling of the hard emission (see Fig. \ref{fig:spectra} and Table \ref{bestfits}).}
\label{fig:podbbfits}
\end{figure}

\begin{figure}
\centerline{\epsfig{file= 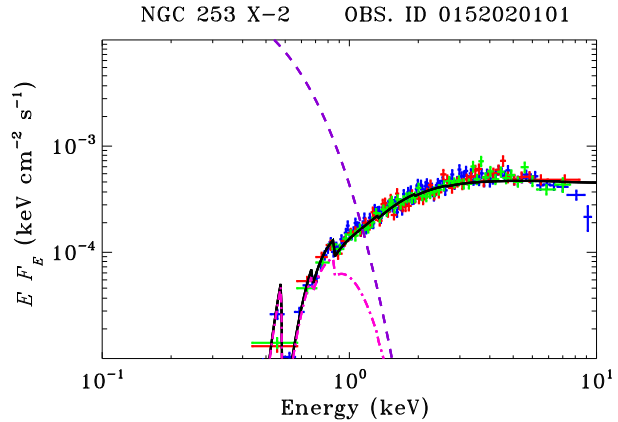,width=8.0cm}}
\caption{ $EF_E$ spectra of NGC 253 X-2 fitted with an absorbed {\sc powerlaw} plus cool MCD model. EPIC pn, mos1 and mos2 data are shown in blue, red and green, respectively. The MCD component is shown by dot-dashed line. The dashed line shows this with the effect of absorption removed. A simple absorbed MCD fit with $\Tin \sim 1.5$ keV improves the fit from $\chired \sim 1.17$ to $\chired \sim 1.08$. Improper modelling of the continuum by a power-law leads to almost an order of magnitude increase in the absorption column density, which produces the highly luminous cool, $\sim 0.1$ keV, MCD component.}
\label{fig:spectra}
\end{figure}

\subsection{The soft excess and the nature of ULXs} 
\label{soft_excess}

When fitting the data with the {\sc powerlaw} plus a cool MCD model, the fit improves for many (but not all) observations compared to the single component fits (see Table \ref{bestfits}). The $\Lsoft-\Tin$ relation for this soft excess is shown in Fig. \ref{fig:podbbfits}. We find that the intrinsic luminosity of the MCD component {\em roughly} scales as $\Lsoft \propto \Tin^{-3.5}$. The best fitting relation derived using the fitexy routine \citep{Press92} is $\Lsoft \propto \Tin^{-3.5 \pm 0.2}$ with $\chi_{\rm{red}}^2 \sim 1.09$. However, this result should be taken with some caution and there are several points to be considered. 
\begin{itemize} 
\item[(i)] The highly luminous low-temperature excesses ($\Lsoft \gtrsim 10^{40}\, \ergs$, $\Tin \sim 0.1$ keV) of NGC 253 X-2 do not seem to be real features, but rather artifacts of improper modelling of the harder emission with a power-law. This results in a rise in the modelled absorption column, which then causes the increase in the estimated intrinsic luminosity of the soft excess (see Fig. \ref{fig:spectra}). We therefore did not include the soft excesses of NGC 253 X-2 in deriving the best fitting $\Lsoft - \Tin$ relation.

\item[(ii)] As in the {\sc powerlaw} fits, $\NH$ and $\Gamma$ are still correlated for NGC 1313 ULX-1 and NGC 5204 ULX-1, but no longer for Holmberg IX ULX-1. As discussed earlier (see subsection \ref{po_type}), there is no clear evidence whether this correlation has a physical origin, or the power-law component has a low-energy break, what one would expect at about $\Tin$, if it arises from Comptonization of the disc photons in a corona. If such break would exist in the power-law component (note that $\Gamma \gtrsim 2$ for NGC 1313 ULX-1 and NGC 5204 ULX-1), then it would necessarily lead to both lower absorption columns and higher {\it absorbed} MCD component luminosities to compensate the `loss' of flux at low energies. But on the other hand, the decrease in $\NH$ could then ultimately lead to lower {\it deabsorbed} luminosities $\Lsoft$.
 
\item[(iii)] If one interprets the spectra in terms of an outflow model \citep{PLF07}, one would then expect an increase of $\NH$ with the luminosity because the outflow would increase the column depth. Also, $\NH$ could be then inversely proportional to $\Tin$, as it should decrease with increasing luminosity. 

\item[(iv)] The host galaxy distances have some uncertainties, which can lead to systematic errors in the derived luminosities.

\item[(v)] The MCD component appears at low energies, where the calibration of {\it XMM-Newton} and {\it Chandra} are the least accurate. Also, at these energies the correct estimation of the absorption column makes the most difference in the derived parameter values of the MCD component. 

\item[(vi)] In addition to these complications, the selection of the lower energy bound in the spectral fitting also has some effect. For example, there are some discrepancies between our results of NGC 5204 ULX-1 and those of \citet{FK09}, who used $0.2$ keV as a lower energy bound in their fitting.
\end{itemize}
All these factors cast a doubt on the derived best fitting relation $\Lsoft \propto \Tin^{-3.5 \pm 0.2}$, but the overall trend of cooler discs with higher luminosities at $\Lsoft \lesssim 10^{40}\, \ergs$ and $\Tin \gtrsim 0.1$ keV seems to be real. However, the observed trend (whether or not it has a power-law index $-3.5$ or something close) is not consistent with the interpretation of an accretion disc around an IMBH. If the observed soft excess does not originate from a disc around an IMBH, what is its origin? One possible explanation could be that the thermal emission arises from the spherization radius of a super-Eddington disc with an outflow \citep{PLF07}. As observed, this model predicts that the thermal emission becomes cooler with increasing luminosity. Although the observed trend is too steep for the predicted relation of this model \citep[][ eqs. 19, 36--38]{PLF07}, the apparent luminosity of this component might be overestimated because the outflowing wind causes some of the emission to be beamed. Indeed, \citet{KP02} showed that at low inclination angles an outflow causes the observed luminosity to be 
\be \label{Lking1}
L_{\rm obs} = 2.3\times 10^{44}\left(\frac{T}{0.1 {\rm keV}}\right)^{-4} \frac{l^2}{pbr^2}\, \ergs,
\ee
where the luminosity is scaled to the Eddington value $l = L/\Ledd$, $p$ is a factor allowing for deviation from the spherical symmetry, $r$ is the radius scaled to the Schwarzschild radius $r = R/R_{\rm s}$ ($R_{\rm s} = 2GM / c^2$) and $b$ is the beaming factor. Assuming that $\Lsoft = L_{\rm obs}$, \citet{K09} argued that the beaming factor has to scale as $b \propto \mdot^{-2}$, where the accretion rate is scaled to its Eddington value $\mdot = \Mdot / \Mdotedd$, to satisfy the observed relation $\Lsoft \sim 7 \times 10^{40} \Tin^{-4}$ based on the results of \citet{KP08}. Using this scaling, \citet{K09} estimated the mass of the accreting black hole 
\be \label{ULXmass}
\frac{M}{\msun} = x \frac{4500}{\mdot^2 (1+ {\rm ln}\,\mdot)} \frac{\Lsoft}{10^{40} \, \ergs},
\ee
where the factor $x$ is the order of unity \citep[see][ eq. 7]{K09}. In our view, however, the estimation of the black hole mass given by equation (\ref{ULXmass}) must be taken with caution, because of the possible systematic uncertainties in the derived best fitting parameters of the MCD component. Never the less, we conclude that our observational data is consistent with the idea that the most luminous sources in our sample are $\sim 10 \msun$ black holes accreting at mildly super-Eddington rates $\mdot \sim 10$ and viewed at low inclinations.

The ULXs that have this soft excess in our sample have a power-law type spectra. This spectrum most likely arises from Comptonization of the disc photons in a corona surrounding the central parts (within the spherization radius) of the accretion flow \citep{HM93,PS96}. Another possibility for the power-law spectra could be due to overheating of the accreting gas close to accreting black hole \citep{B98}. Therefore, a plausible explanation to both the observed $\Lsoft - \Tin$ relation and power-law type spectra could be that at low inclinations the central Comptonized part of the disc would be visible to the observer at all times and the outflowing wind would cause the Comptonized emission and the emission from the spherization radius to be geometrically beamed, thus resulting in a high $\sim 10^{40} \, \ergs$ apparent luminosity. On the other hand, at higher inclinations the central part of the disc would be obscured by the outflow. This would result in a non-power-law type spectra that we see for NGC 253 X-4 and NGC 253 X-9.

\section{Conclusions}

We have studied spectral variability of 11 ULXs and can draw several conclusions.
 
\begin{itemize}
\item[(i)] The ULXs of this sample can be roughly separated into two types based on their spectra: power-law- and non-power-law type ULXs. All the power-law type ULXs are constantly more luminous than $\sim 3 \times 10^{39} \, \ergs$, i.e. they show permanently super-Eddington luminosities for a $\sim 10 \msun$ black hole.

\item[(ii)] Six out of the seven power-law type ULXs show a $\Lx-\Gamma$ correlation, i.e. the observed spectrum softens as the luminosity increases. Furthermore, four ULXs show also spectral pivoting with pivot energies $\epiv \lesssim 10$ keV. Three ULXs (NGC 1313 ULX-1, Holmberg II ULX-1 and NGC 5204 ULX-1) also show state transitions to a low/soft state.

\item[(iii)] The non-power-law type ULXs can be adequately modelled with a MCD. These sources also follow the behaviour of known BHBs in the $L-T$ diagrams.

\item[(iv)] NGC 1313 ULX-2 is a very interesting source showing a VHS type spectra (similar to XTE J1550-564) at lower luminosities and a non-power-law (thermal) like spectra at high luminosities. It is the only source in this sample showing transitions between a spectral state that is seen in known BHBs and an ultraluminous spectral state. NGC 1313 ULX-2 is therefore most likely a super-Eddington source. 

\item[(v)] The soft excess seems to roughly follow a trend $\Lsoft \propto \Tin^{-3.5}$, which does not support the idea that the observed soft excess arises from the disc emission around IMBHs. Instead, the soft excess in consistent with beamed disc emission from the spherization radius. 
\end{itemize}

We therefore conclude that the ULXs are most likely powered by super-Eddington accretion onto $\sim 10 \msun$ black holes. The two different types of ULXs result from their different inclination angles. The permanently super-Eddington power-law type ULXs might be viewed at low inclination angles. The low-luminosity ULXs with non-power-law spectra are consistent of being super-Eddington sources viewed at high inclinations (NGC 253 X-4, NGC 253 X-9) or they are sub-Eddington sources (NGC 253 X-2).

\section*{Acknowledgements}

The authors thank Dr. Hua Feng and the anonymous referee for useful comments. 
This work was supported by the Finnish Graduate School in Astronomy and Space Physics (JJEK) and  
the Academy of Finland grant 127512 (JP).
This research has made use of data obtained from the Chandra Data Archive and software provided by the Chandra X-ray Center (CXC) in the application packages CIAO, ChIPS, and Sherpa.
This research was based on observations obtained with XMM-Newton, an ESA science mission with instruments and contributions directly funded by ESA Member States and NASA.


\clearpage

\appendix

\onecolumn

\section{Best-fitting model parameters}


 \renewcommand{\thefootnote}{\alph{footnote}}

\begin{longtable}{c c c c c c c c}
\caption{Best-fitting parameters and absorption corrected luminosity estimates.
The results are given only if $\chi_{\rm{red}}^2 < 2$. 
In the case of the {\sc powerlaw} $+$ {\sc diskbb} model, the results are given if the soft excess {\sc diskbb} component is constrained. 
} \label{bestfits} \\

\hline \hline \\[-2ex]
 Obs.ID & $\NH$ $^{b}$ & $\Gamma$ $^{c}$ & $T_{\rm in}$ $^{d}$ & $L_{\rm power-law}$ $^{e}$ & $L_{\rm diskbb}$ $^{f}$ & $\chi_{\rm red}^2\,|\,$d.o.f. $^{g}$ & $P_{\rm rej}$ $^{h}$ \\[0.5ex] \hline
   \\[-1.8ex]
\endfirsthead

\multicolumn{8}{c}{{\tablename} \thetable{} -- Continued} \\[0.5ex]
\hline \hline \\[-2ex]
 Obs.ID & $\NH$ $^{a}$ & $\Gamma$ $^{b}$ & $T_{\rm{in}}$ $^{c}$ & $L_{\rm{power-law}}$ $^{d}$ & $L_{\rm{diskbb}}$ $^{e}$ & $\chi_{\rm{red}}^2\,|\,$d.o.f. $^{f}$ & $P_{\rm{rej}}$ $^{g}$ \\[0.5ex] \hline
   \\[-1.8ex]
\endhead

\\[0.5ex] \hline
   \\[-1.8ex]
  \multicolumn{8}{l}{{Continued on next page\ldots}}
\endfoot

  \\[-1.8ex] \hline \hline
\endlastfoot


\multicolumn{8}{c}{NGC 253 X-2} \\
  \\[-1.8ex]
    790			& $ 1.4_{-0.3}^{+0.3}$  	& $ ... $ 					& $ 1.28_{-0.09}^{+0.10}$	& $ ... $ 					& $ 0.92_{-0.02}^{+0.02}$  		& $ 1.80\,|\,105$ & 100.0\%\\ 
    			& $ 4.0_{-0.5}^{+0.5}$ 	& $ 2.1_{-0.1}^{+0.1}$	& $ ... $ 					& $ 1.37_{-0.06}^{+0.07}$ 	& $ ... $ 						& $ 1.26\,|\,105$ & 96.2\%\\ 
				& $ 5.7_{-1.5}^{+2.0}$ 	& $ 2.0_{-0.2}^{+0.2}$ 	& $ 0.22_{-0.06}^{+0.10}$ 	& $ 1.3_{-0.2}^{+0.2}$ 	& $ 1.1_{-0.8}^{+3.8}$  		& $ 1.19\,|\,103$ &  90.3\%\\

    0110900101	& $ 1.7_{-0.2}^{+0.2}$ 	& $ ... $ 					& $ 1.64_{-0.06}^{+0.07}$ 	& $ ... $ 					& $ 2.38_{-0.04}^{+0.04}$ 		& $ 1.09\,|\,348$ &  88.1\%\\ 
    			& $ 3.7_{-0.3}^{+0.3}$ 	& $ 1.76_{-0.05}^{+0.05}$ 	& $ ... $ 					& $ 3.0_{-0.1}^{+0.1}$ 	& $ ... $ 						& $ 1.18\,|\,348$ &  98.7\%\\ 
				& $ 8.7_{-1.2}^{+1.2}$ 	& $ 2.1_{-0.1}^{+0.1}$ 	& $ 0.10_{-0.01}^{+0.01}$ 	& $ 3.8_{-0.3}^{+0.3}$ 	& $ 92_{-56}^{+142}$  	& $ 1.05\,|\,346$ &  76.3\%\\ 

    0125960101	& $ 1.56_{-0.11}^{+0.11}$ 	& $ ... $ 					& $ 1.52_{-0.04}^{+0.04}$	& $ ... $ 					& $ 1.87_{-0.02}^{+0.02}$ 		& $ 0.95\,|\,583$ &  18.2\%\\ 
     			& $ 3.7_{-0.2}^{+0.2}$ 	& $ 1.89_{-0.04}^{+0.04}$ 	& $ ... $ 				 	& $ 2.37_{-0.05}^{+0.05}$ 	& $ ... $ 						& $ 1.24\,|\,583$ & 100.0\%\\ 
    			& $ 9.3_{-0.8}^{+0.8}$ 	& $ 2.26_{-0.06}^{+0.06}$ 	& $ 0.09_{-0.01}^{+0.01}$ 	& $ 3.4_{-0.2}^{+0.3}$	& $ 202_{-89}^{+195}$ 	& $ 1.01\,|\,581$ &  58.9\%\\ 

    0125960201	& $ 1.8_{-0.3}^{+0.3}$ 	& $ ... $ 					& $ 1.83_{-0.11}^{+0.12}$	& $ ... $ 					& $ 2.68_{-0.07}^{+0.07}$ 		& $ 0.97\,|\,160$ &  41.5\%\\ 
     			& $ 3.7_{-0.5}^{+0.5}$ 	& $ 1.66_{-0.08}^{+0.08}$ 	& $ ... $ 					& $ 3.15_{-0.13}^{+0.14}$ 	& $ ... $ 						& $ 1.25\,|\,160$ &  98.1\%\\ 
    			& $ 8.9_{-1.7}^{+1.7}$ 	& $ 2.0_{-0.1}^{+0.1}$ 	& $ 0.09_{-0.01}^{+0.01}$ 	& $ 4.0_{-0.4}^{+0.5}$ 	& $ 189_{-138}^{+519}$ & $ 1.04\,|\,158$ &  63.9\%\\ 

    0152020101	& $ 1.65_{-0.08}^{+0.08}$ 	& $ ... $ 					& $ 1.66_{-0.03}^{+0.03}$	& $ ... $ 					& $ 1.87_{-0.01}^{+0.02}$ 		& $ 1.08\,|\,934$ &  95.8\%\\ 
     			& $ 3.7_{-0.14}^{+0.14}$ 	& $ 1.80_{-0.03}^{+0.03}$ 	& $ ... $ 					& $ 2.28_{-0.03}^{+0.03}$ 	& $ ... $ 						& $ 1.32\,|\,934$ & 100.0\%\\ 
    			& $ 8.2_{-0.6}^{+0.6}$ 	& $ 2.08_{-0.05}^{+0.05}$ 	& $ 0.10_{-0.01}^{+0.01}$ 	& $ 2.9_{-0.1}^{+0.1}$ 	& $ 55_{-21}^{+ 36}$  	& $ 1.17\,|\,932$ & 100.0\%\\ 

    0304850901	& $ 2.0_{-0.3}^{+0.3}$ 	& $ ... $ 					& $ 1.53_{-0.08}^{+0.09}$ 	& $ ... $ 					& $ 1.55_{-0.02}^{+0.03}$ 		& $ 1.02\,|\,166$ &  59.1\%\\ 
     			& $ 4.3_{-0.5}^{+0.5}$ 	& $ 1.89_{-0.08}^{+0.09}$ 	& $ ... $ 					& $ 2.05_{-0.09}^{+0.11}$ 	& $ ... $ 						& $ 1.22\,|\,166$ &  97.0\%\\ 
    			& $ 9.2_{-1.9}^{+2.0}$ 	& $ 2.21_{-0.14}^{+0.15}$ 	& $ 0.09_{-0.01}^{+0.01}$ 	& $ 2.8_{-0.4}^{+0.5}$ 	& $ 100_{-76}^{+316}$ 	& $ 1.10\,|\,164$ &  80.8\%\\ 

    0304851001	& $ 1.5_{-0.2}^{+0.2}$ 	& $ ... $ 					& $ 1.59_{-0.09}^{+0.09}$ 	& $ ... $ 					& $ 1.52_{-0.03}^{+0.03}$ 		& $ 0.79\,|\,168$ &   1.9\%\\ 
     			& $ 3.5_{-0.4}^{+0.4}$ 	& $ 1.82_{-0.08}^{+0.08}$ 	& $ ... $ 					& $ 1.91_{-0.07}^{+0.08}$ 	& $ ... $ 						& $ 0.89\,|\,168$ &  16.8\%\\ 
    			& $ 8.0_{-1.7}^{+1.7}$ 	& $ 2.09_{-0.13}^{+0.13}$ 	& $ 0.10_{-0.01}^{+0.02}$ 	& $ 2.4_{-0.3}^{+0.3}$ 	& $ 46_{-35}^{+132}$ 	& $ 0.80\,|\,166$ &   2.9\%\\ 

    0304851101	& $ 1.3_{-0.2}^{+0.3}$ 	& $ ... $ 					& $ 1.54_{-0.09}^{+0.09}$ 	& $ ... $ 					& $ 1.33_{-0.03}^{+0.03}$ 		& $ 0.96\,|\,168$ &  36.4\%\\ 
     			& $ 3.2_{-0.4}^{+0.4}$ 	& $ 1.79_{-0.08}^{+0.08}$ 	& $ ... $ 					& $ 1.67_{-0.08}^{+0.09}$ 	& $ ... $ 						& $ 1.21\,|\,168$ &  96.9\%\\ 
    			& $ 9.1_{-1.6}^{+1.6}$ 	& $ 2.20_{-0.14}^{+0.14}$ 	& $ 0.09_{-0.01}^{+0.01}$ 	& $ 2.4_{-0.3}^{+0.4}$ 	& $ 142_{-104}^{+379}$ & $ 1.03\,|\,166$ &  62.6\%\\ 

    0304851201	& $ 1.5_{-0.2}^{+0.2}$ 	& $ ... $ 					& $ 1.51_{-0.07}^{+0.07}$ 	& $ ... $ 					& $ 1.45_{-0.02}^{+0.02}$ 		& $ 1.12\,|\,247$ &  89.8\%\\
    			& $ 3.5_{-0.3}^{+0.3}$ 	& $ 1.84_{-0.07}^{+0.07}$ 	& $ ... $ 					& $ 1.85_{-0.06}^{+0.07}$ 	& $ ... $ 						& $ 1.39\,|\,247$ & 100.0\%\\
    			& $ 8.1_{-1.2}^{+1.2}$ 	& $ 2.18_{-0.11}^{+0.11}$ 	& $ 0.09_{-0.01}^{+0.01}$ 	& $ 2.5_{-0.2}^{+0.3}$ 	& $ 99_{-64}^{+191}$  	& $ 1.23\,|\,245$ &  99.1\%\\ 
  \\[-1.8ex]
\multicolumn{8}{c}{NGC 253 X-4} \\
  \\[-1.8ex]
    790			& $ 2.3_{-0.5}^{+0.5}$ & $ ... $ 					& $ 1.12_{-0.08}^{+0.08}$ 	& $ ... $ 				  	& $ 0.50_{-0.02}^{+0.02}$ 		& $ 0.85\,|\,62$ &  21.4\%\\ 
     			& $ 5.5_{-0.7}^{+0.8}$ & $ 2.37_{-0.14}^{+0.14}$ 	& $ ... $ 					& $ 0.83_{-0.08}^{+0.10}$ 	& $ ... $ 						& $ 1.08\,|\,62$ &  69.5\%\\ 

    0110900101	& $ 2.0_{-0.5}^{+0.5}$ & $ ... $ 					& $ 1.19_{-0.09}^{+0.10}$ 	& $ ... $ 				  	& $ 0.42_{-0.01}^{+0.01}$ 		& $ 0.70\,|\,88$ &  1.5\%\\ 
     			& $ 5.0_{-0.8}^{+0.8}$ & $ 2.21_{-0.14}^{+0.15}$ 	& $ ... $ 					& $ 0.69_{-0.07}^{+0.08}$ 	& $ ... $ 						& $ 0.80\,|\,88$ &  8.6\%\\ 

    0125960101	& $ 3.2_{-0.3}^{+0.3}$ & $ ... $ 					& $ 1.00_{-0.04}^{+0.04}$ 	& $ ... $ 				  	& $ 0.53_{-0.01}^{+0.01}$ 		& $ 1.07\,|\,260$ & 78.7\%\\ 
     			& $ 7.1_{-0.5}^{+0.6}$ & $ 2.67_{-0.09}^{+0.09}$ 	& $ ... $ 					& $ 1.1_{-0.1}^{+0.1}$ 	& $ ... $ 						& $ 1.18\,|\,260$ & 97.4\%\\ 

    0152020101	& $ 3.2_{-0.2}^{+0.2}$ & $ ... $ 					& $ 1.07_{-0.03}^{+0.03}$ 	& $ ... $ 				  	& $ 0.57_{-0.01}^{+0.01}$ 		& $ 0.95\,|\,497$ & 22.8\%\\ 
     			& $ 7.0_{-0.4}^{+0.4}$ & $ 2.54_{-0.06}^{+0.07}$ 	& $ ... $ 					& $ 1.08_{-0.06}^{+0.07}$ 	& $ ... $ 						& $ 1.20\,|\,497$ & 99.8\%\\ 

    3931		& $ 2.2_{-0.3}^{+0.3}$ & $ ... $ 					& $ 1.18_{-0.06}^{+0.06}$ 	& $ ... $ 				  	& $ 0.53_{-0.01}^{+0.01}$ 		& $ 1.05\,|\,133$ & 67.0\%\\ 
     			& $ 4.9_{-0.4}^{+0.5}$ & $ 2.2_{-0.1}^{+0.1}$ 	& $ ... $ 					& $ 0.81_{-0.04}^{+0.05}$ 	& $ ... $ 						& $ 1.26\,|\,133$ & 97.8\%\\ 

    0304850901	& $ 2.2_{-0.5}^{+0.6}$ & $ ... $ 					& $ 1.15_{-0.10}^{+0.11}$ 	& $ ... $ 				  	& $ 0.51_{-0.02}^{+0.02}$ 		& $ 0.73\,|\,75$ &   4.0\%\\ 
     			& $ 5.3_{-0.9}^{+1.0}$ & $ 2.3_{-0.2}^{+0.2}$ 	& $ ... $ 					& $ 0.83_{-0.09}^{+0.12}$ 	& $ ... $ 						& $ 0.83\,|\,75$ &  14.4\%\\ 

    0304851001	& $ 3.4_{-0.7}^{+0.8}$ & $ ... $ 					& $ 1.02_{-0.09}^{+0.10}$ 	& $ ... $ 				  	& $ 0.53_{-0.02}^{+0.03}$ 		& $ 1.15\,|\,76$ &  81.9\%\\
     			& $ 7.2_{-1.1}^{+1.3}$ & $ 2.5_{-0.2}^{+0.2}$ 	& $ ... $ 					& $ 1.0_{-0.2}^{+0.2}$ 	& $ ... $ 						& $ 1.14\,|\,76$ &  80.4\%\\ 
  \\[-1.8ex]
\multicolumn{8}{c}{NGC 253 X-9} \\
  \\[-1.8ex]
    0110900101	& $ 1.6_{-0.3}^{+0.4}$ & $ 2.4_{-0.2}^{+0.2}$ 	& $ ... $ 					& $ 0.41_{-0.04}^{+0.04}$ 	& $ ... $ 						& $ 1.25\,|\,104$ &  95.5\%\\ 

    0125960101	& $ 0.37_{-0.17}^{+0.18}$ & $ ... $ 					& $ 1.33_{-0.07}^{+0.07}$ 	& $ ... $ 				  	& $ 1.13_{-0.02}^{+0.02}$ 		& $ 0.99\,|\,173$ &  46.1\%\\ 
     			& $ 2.2_{-0.3}^{+0.3}$ & $ 1.93_{-0.07}^{+0.08}$ 	& $ ... $ 					& $ 1.53_{-0.05}^{+0.06}$ 	& $ ... $ 						& $ 1.24\,|\,173$ &  98.3\%\\ 

    0152020101	& $ 0.41_{-0.06}^{+0.06}$ & $ ... $ 					& $ 1.19_{-0.03}^{+0.03}$ 	& $ ... $ 				  	& $ 1.14_{-0.01}^{+0.01}$ 		& $ 1.14\,|\,821$ &  99.7\%\\ 
     			& $ 2.4_{-0.1}^{+0.1}$ & $ 2.12_{-0.03}^{+0.03}$ 	& $ ... $ 					& $ 1.63_{-0.03}^{+0.03}$ 	& $ ... $ 						& $ 1.20\,|\,821$ & 100.0\%\\ 

    3931		& $ 0.3_{-0.1}^{+0.1}$ & $ ... $ 					& $ 1.43_{-0.07}^{+0.07}$ 	& $ ... $ 				  	& $ 0.84_{-0.02}^{+0.02}$ 		& $ 1.31\,|\,180$ &  99.7\%\\ 
     			& $ 1.8_{-0.2}^{+0.2}$ & $ 1.74_{-0.06}^{+0.06}$ 	& $ ... $ 					& $ 1.11_{-0.01}^{+0.01}$ 	& $ ... $ 						& $ 1.07\,|\,180$ &  74.2\%\\ 

    0304850901	& $ 0.17_{-0.03}^{+0.25}$ & $ ... $ 					& $ 1.29_{-0.10}^{+0.10}$ 	& $ ... $ 				  	& $ 0.50_{-0.01}^{+0.02}$ 		& $ 1.03\,|\,90$ &  58.9\%\\ 
     			& $ 1.8_{-0.4}^{+0.4}$ & $ 1.93_{-0.1}^{+0.1}$ 	& $ ... $ 					& $ 0.66_{-0.03}^{+0.04}$ 	& $ ... $ 						& $ 1.25\,|\,90$ &  94.6\%\\ 

    0304851001	& $ 0.14_{-0.01}^{+0.22}$ & $ ... $ 					& $ 1.36_{-0.11}^{+0.10}$ 	& $ ... $ 				  	& $ 0.56_{-0.02}^{+0.02}$ 		& $ 1.06\,|\,100$ &  69.0\%\\ 
     			& $ 1.8_{-0.3}^{+0.4}$ & $ 1.88_{-0.1}^{+0.1}$ 	& $ ... $ 					& $ 0.75_{-0.03}^{+0.03}$ 	& $ ... $ 						& $ 1.00\,|\,100$ &  50.7\%\\ 

    0304851201	& $ 0.4_{-0.2}^{+0.2}$ & $ ... $ 					& $ 1.32_{-0.08}^{+0.09}$ 	& $ ... $ 				  	& $ 0.61_{-0.01}^{+0.01}$ 		& $ 1.02\,|\,156$ &  57.4\%\\
     			& $ 2.2_{-0.3}^{+0.3}$ & $ 1.93_{-0.1}^{+0.1}$ 	& $ ... $ 					& $ 0.83_{-0.04}^{+0.04}$ 	& $ ... $ 						& $ 1.14\,|\,156$ &  89.3\%\\ 
  \\[-1.8ex]
\multicolumn{8}{c}{NGC 1313 ULX-1} \\
  \\[-1.8ex]
    0106860101	& $ 1.70_{-0.09}^{+0.09}$ & $ 1.86_{-0.03}^{+0.03}$ 	& $ ... $ 					& $ 5.52_{-0.06}^{+0.07}$ 	& $ ... $ 						& $ 1.16\,|\,810$ &  99.9\%\\ 
				& $ 2.7_{-0.3}^{+0.4}$ & $ 1.71_{-0.05}^{+0.05}$ 	& $ 0.21_{-0.02}^{+0.03}$ 	& $ 5.3_{-0.2}^{+0.2}$ 	& $ 3.4_{-1.0}^{+1.6}$ 		& $ 0.99\,|\, 808$ &  42.9\%\\ 

    0150280301	& $ 3.4_{-0.2}^{+0.2}$ & $ 2.42_{-0.07}^{+0.07}$ 	& $ ... $ 					& $ 15.2_{-0.6}^{+0.6}$ 	& $ ... $ 						& $ 0.99\,|\,261$ &  44.8\%\\ 
    			& $ 4.0_{-0.6}^{+0.9}$ & $ 2.36_{-0.10}^{+0.10}$ 	& $ 0.18_{-0.04}^{+0.08}$ 	& $ 14.7_{-1.3}^{+1.4}$ 	& $ 7_{-5}^{+19}$ 		& $ 0.96\,|\, 259$ &  34.8\%\\ 

    0150280401	& $ 3.7_{-0.4}^{+0.4}$ & $ 2.4_{-0.1}^{+0.1}$ 	& $ ... $ 					& $ 12.5_{-0.8}^{+1.0}$ 	& $ ... $ 						& $ 0.95\,|\,88$ &  39.8\%\\ 
    			& $ 4.4_{-1.0}^{+1.9}$ & $ 2.3_{-0.2}^{+0.2}$ 	& $ 0.18_{-0.06}^{+0.18}$ 	& $ 12.1_{-2.0}^{+2.4}$ 	& $ 7_{-5}^{+19}$ 		& $ 0.94\,|\,  86$ &  37.0\%\\ 

    0150280601	& $ 3.0_{-0.2}^{+0.2}$ & $ 2.31_{-0.08}^{+0.08}$ 	& $ ... $ 					& $ 10.8_{-0.4}^{+0.5}$ 	& $ ... $ 						& $ 0.99\,|\,195$ &  49.1\%\\ 
    			& $ 3.0_{-0.6}^{+0.6}$ & $ 2.1_{-0.3}^{+0.2}$ 	& $ 0.29_{-0.09}^{+0.12}$ 	& $ 9.0_{-1.8}^{+1.4}$ 	& $ 2.8_{-1.5}^{+2.7}$ 		& $ 0.96\,|\, 193$ &  34.4\%\\ 

    0150281101	& $ 3.2_{-0.2}^{+0.2}$ & $ 2.35_{-0.07}^{+0.07}$ 	& $ ... $ 					& $ 12.0_{-0.6}^{+0.6}$ 	& $ ... $ 						& $ 0.92\,|\,262$ &  19.7\%\\ 

    0205230201	& $ 2.3_{-0.3}^{+0.4}$ & $ 2.06_{-0.11}^{+0.12}$ 	& $ ... $ 					& $ 6.1_{-0.3}^{+0.3}$ 	& $ ... $ 						& $ 1.16\,|\,104$ &  87.8\%\\ 
    			& $ 4.7_{-1.7}^{+1.8}$ & $ 2.0_{-0.2}^{+0.2}$ 	& $ 0.15_{-0.03}^{+0.08}$ 	& $ 6.3_{-0.9}^{+1.1}$ 	& $ 18_{-15}^{+86}$ 	& $ 1.04\,|\, 102$ &  63.7\%\\ 

    0205230301	& $ 3.4_{-0.2}^{+0.2}$ & $ 2.31_{-0.05}^{+0.06}$ 	& $ ... $ 					& $ 17.8_{-0.6}^{+0.7}$ 	& $ ... $ 						& $ 0.98\,|\,302$ &  41.3\%\\ 

    0205230401	& $ 3.9_{-0.2}^{+0.3}$ & $ 3.3_{-0.1}^{+0.1}$ 	& $ ... $ 					& $ 9.4_{-0.9}^{+1.1}$ 	& $ ... $ 						& $ 1.18\,|\,238$ &  97.0\%\\ 
    			& $ 4.6_{-0.6}^{+0.8}$ & $ 3.0_{-0.2}^{+0.2}$ 	& $ 0.16_{-0.02}^{+0.03}$ 	& $ 6.9_{-1.3}^{+1.4}$ 	& $ 14_{-7}^{+20}$ 	& $ 1.03\,|\, 236$ &  63.7\%\\ 

    0205230501	& $ 2.2_{-0.2}^{+0.2}$ & $ 2.04_{-0.07}^{+0.08}$ 	& $ ... $ 					& $ 7.5_{-0.2}^{+0.2}$ 	& $ ... $ 						& $ 1.10\,|\,212$ &  85.9\%\\ 
    			& $ 2.4_{-0.5}^{+0.6}$ & $ 1.7_{-0.2}^{+0.2}$ 	& $ 0.30_{-0.07}^{+0.08}$ 	& $ 6.3_{-0.6}^{+0.6}$ 	& $ 2.9_{-0.9}^{+1.6}$ 		& $ 0.97\,|\, 210$ &  39.8\%\\ 

   0205230601	& $ 1.9_{-0.2}^{+0.2}$ & $ 1.92_{-0.06}^{+0.06}$ 	& $ ... $ 					& $ 6.2_{-0.1}^{+0.1}$ 	& $ ... $ 						& $ 1.18\,|\,331$ &  98.6\%\\ 
    			& $ 3.0_{-0.6}^{+0.7}$ & $ 1.74_{-0.10}^{+0.10}$ 	& $ 0.21_{-0.03}^{+0.04}$ 	& $ 5.9_{-0.3}^{+0.3}$ 	& $ 4.7_{-2.0}^{+4.3}$ 		& $ 1.03\,|\, 329$ &  63.9\%\\ 

   0405090101	& $ 1.79_{-0.04}^{+0.04}$ & $ 1.86_{-0.02}^{+0.02}$ 	& $ ... $ 					& $ 5.37_{-0.03}^{+0.03}$ 	& $ ... $ 						& $ 1.34\,|\,1514$ & 100.0\%\\ 
    			& $ 2.9_{-0.2}^{+0.2}$ & $ 1.73_{-0.03}^{+0.03}$ 	& $ 0.20_{-0.01}^{+0.01}$ 	& $ 5.2_{-0.1}^{+0.1}$ 	& $ 3.8_{-0.7}^{+0.8}$ 		& $ 1.01\,|\,1512$ &  58.8\%\\ 

  \\[-1.8ex]

\multicolumn{8}{c}{NGC 1313 ULX-2} \\
  \\[-1.8ex]
    0106860101	& $ 2.5_{-0.2}^{+0.2}$ & $ 2.4_{-0.1}^{+0.1}$ 	& $ ... $ 					& $ 2.5_{-0.1}^{+0.1}$ 	& $ ... $ 						& $ 0.98\,|\,172$ &  44.1\%\\ 
   				& $ 2.5_{-0.5}^{+0.6}$ & $ 2.1_{-0.3}^{+0.2}$ 	& $ 0.27_{-0.08}^{+0.09}$ 	& $ 2.0_{-0.4}^{+0.3}$ 	& $ 0.9_{-0.4}^{+0.7}$ 		& $ 0.91\,|\,170$ &  21.5\%\\ 

    0150280101	& $ 0.9_{-0.3}^{+0.3}$ & $ ... $ 					& $ 1.4_{-0.1}^{+0.1}$ 	& $ ... $ 				  	& $ 4.5_{-0.1}^{+0.1}$ 		& $ 1.03\,|\,78$ &  60.4\%\\ 
     			& $ 2.6_{-0.5}^{+0.5}$ & $ 1.8_{-0.1}^{+0.1}$ 	& $ ... $ 					& $ 6.2_{-0.4}^{+0.4}$ 	& $ ... $ 						& $ 1.10\,|\,78$ &  74.0\%\\ 

    0150280301	& $ 0.86_{-0.10}^{+0.10}$ & $ ... $ 					& $ 1.63_{-0.05}^{+0.06}$ 	& $ ... $ 				  	& $ 6.1_{-0.1}^{+0.1}$ 		& $ 1.18\,|\,450$ &  99.4\%\\ 
     			& $ 2.6_{-0.2}^{+0.2}$ & $ 1.76_{-0.04}^{+0.04}$ 	& $ ... $ 					& $ 7.5_{-0.1}^{+0.1}$ 	& $ ... $ 						& $ 1.07\,|\,450$ &  84.8\%\\ 
    			& $ 4.4_{-1.0}^{+0.9}$ & $ 1.89_{-0.08}^{+0.07}$ 	& $ 0.12_{-0.01}^{+0.02}$ 	& $ 8.2_{-0.4}^{+0.5}$ 	& $ 12_{-9}^{+21}$ 	& $ 1.05\,|\,448$ &  78.9\%\\ 

    0150280401	& $ 0.86_{-0.13}^{+0.14}$ & $ ... $ 					& $ 1.69_{-0.08}^{+0.08}$ 	& $ ... $ 				  	& $ 7.0_{-0.2}^{+0.2}$ 		& $ 1.04\,|\,284$ &  69.4\%\\ 
     			& $ 2.5_{-0.2}^{+0.2}$ & $ 1.71_{-0.05}^{+0.06}$ 	& $ ... $ 					& $ 8.5_{-0.2}^{+0.2}$ 	& $ ... $ 						& $ 0.95\,|\,284$ &  27.4\%\\ 

    0150280501	& $ 2.3_{-0.3}^{+0.3}$ & $ 2.1_{-0.1}^{+0.1}$ 	& $ ... $ 					& $ 3.9_{-0.3}^{+0.3}$ 	& $ ... $ 						& $ 1.14\,|\,133$ &  87.0\%\\ 

    0150280601	& $ 0.52_{-0.11}^{+0.14}$ & $ ... $ 					& $ 0.98_{-0.04}^{+0.05}$ 	& $ ... $ 				  	& $ 2.04_{-0.03}^{+0.03}$ 		& $ 1.15\,|\,232$ &  93.9\%\\ 
     			& $ 2.8_{-0.2}^{+0.2}$ & $ 2.4_{-0.1}^{+0.1}$ 	& $ ... $ 					& $ 3.3_{-0.2}^{+0.2}$ 	& $ ... $ 						& $ 0.85\,|\,232$ &   4.6\%\\ 

    0150281101	& $ 2.4_{-0.3}^{+0.4}$ & $ 2.3_{-0.1}^{+0.1}$ 	& $ ... $ 					& $ 2.7_{-0.2}^{+0.2}$ 	& $ ... $ 						& $ 0.97\,|\,116$ &  43.6\%\\ 

    0205230201	& $ 0.6_{-0.2}^{+0.4}$ & $ ... $ 					& $ 0.88_{-0.07}^{+0.08}$ 	& $ ... $ 				  	& $ 1.46_{-0.03}^{+0.05}$ 		& $ 1.27\,|\,58$ &  91.7\%\\ 
     			& $ 3.2_{-0.5}^{+0.6}$ & $ 2.6_{-0.2}^{+0.2}$ 	& $ ... $ 					& $ 2.7_{-0.3}^{+0.4}$ 	& $ ... $ 						& $ 1.17\,|\,58$ &  82.3\%\\ 

    0205230301	& $ 1.1_{-0.1}^{+0.1}$ & $ ... $ 					& $ 1.62_{-0.05}^{+0.05}$ 	& $ ... $ 				  	& $ 7.4_{-0.1}^{+0.1}$ 		& $ 1.11\,|\,580$ &  96.1\%\\ 
     			& $ 3.0_{-0.1}^{+0.2}$ & $ 1.81_{-0.04}^{+0.04}$ 	& $ ... $ 					& $ 9.2_{-0.2}^{+0.2}$ 	& $ ... $ 						& $ 1.05\,|\,580$ &  82.4\%\\ 
    			& $ 4.9_{-0.9}^{+0.8}$ & $ 1.92_{-0.06}^{+0.06}$ 	& $ 0.12_{-0.01}^{+0.01}$ 	& $ 10.0_{-0.5}^{+0.5}$ 	& $ 18_{-12}^{+24}$ 	& $ 1.04\,|\,578$ &  74.8\%\\ 

   0205230401	& $ 1.7_{-0.3}^{+0.3}$ & $ 2.2_{-0.1}^{+0.1}$ 	& $ ... $ 					& $ 1.9_{-0.1}^{+0.1}$ 	& $ ... $ 						& $ 0.93\,|\,135$ &  29.6\%\\ 
    			& $ 2.0_{-0.8}^{+1.1}$ & $ 1.9_{-0.3}^{+0.2}$ 	& $ 0.25_{-0.08}^{+0.13}$	& $ 1.6_{-0.3}^{+0.3}$ 	& $ 0.8_{-0.4}^{+1.8}$ 		& $ 0.87\,|\,133$ &  13.5\%\\ 

   0205230501	& $ 2.5_{-0.2}^{+0.2}$ & $ 2.36_{-0.07}^{+0.07}$ 	& $ ... $ 					& $ 2.45_{-0.1}^{+0.1}$ 	& $ ... $ 						& $ 1.05\,|\,270$ &  70.9\%\\ 
    			& $ 2.8_{-0.5}^{+0.7}$ & $ 2.2_{-0.2}^{+0.1}$ 	& $ 0.22_{-0.05}^{+0.08}$ 	& $ 2.1_{-0.3}^{+0.2}$ 	& $ 1.2_{-0.6}^{+1.9}$ 		& $ 0.98\,|\,268$ &  43.0\%\\ 

   0205230601	& $ 1.05_{-0.1}^{+0.1}$ & $ ... $ 					& $ 1.73_{-0.05}^{+0.06}$ 	& $ ... $ 				  	& $ 7.0_{-0.1}^{+0.1}$ 		& $ 1.08\,|\,536$ &  90.4\%\\ 
    			& $ 2.8_{-0.2}^{+0.2}$ & $ 1.71_{-0.04}^{+0.04}$ 	& $ ... $ 					& $ 8.4_{-0.1}^{+0.1}$ 	& $ ... $ 						& $ 1.07\,|\,536$ &  88.0\%\\ 
    			& $ 5.4_{-0.9}^{+0.8}$ & $ 1.87_{-0.07}^{+0.06}$ 	& $ 0.12_{-0.01}^{+0.01}$ 	& $ 9.4_{-0.4}^{+0.5}$ 	& $ 33_{-19}^{+37}$ 	& $ 1.03\,|\,534$ &  71.7\%\\ 

   0301860101	& $ 1.2_{-0.1}^{+0.1}$ & $ ... $ 					& $ 1.59_{-0.04}^{+0.04}$ 	& $ ... $ 				  	& $ 6.82_{-0.07}^{+0.07}$ 		& $ 1.07\,|\,712$ &  90.5\%\\ 
    			& $ 3.12_{-0.13}^{+0.14}$ & $ 1.83_{-0.03}^{+0.03}$ 	& $ ... $ 					& $ 8.5_{-0.1}^{+0.1}$ 	& $ ... $ 						& $ 1.15\,|\,712$ &  99.7\%\\ 
    			& $ 5.6_{-0.7}^{+0.6}$ & $ 2.00_{-0.05}^{+0.05}$ 	& $ 0.11_{-0.01}^{+0.01}$ 	& $ 9.7_{-0.4}^{+0.4}$ 	& $ 38_{-19}^{+32}$ 	& $ 1.10\,|\,710$ &  96.1\%\\ 

   0405090101	& $ 1.03_{-0.04}^{+0.05}$ & $ ... $ 					& $ 1.57_{-0.02}^{+0.03}$ 	& $ ... $ 				  	& $ 6.18_{-0.04}^{+0.04}$ 		& $ 1.32\,|\,891$ & 100.0\%\\ 
    			& $ 2.87_{-0.07}^{+0.07}$ & $ 1.85_{-0.02}^{+0.02}$ 	& $ ... $ 					& $ 7.69_{-0.06}^{+0.06}$ 	& $ ... $ 						& $ 1.18\,|\,891$ & 100.0\%\\ 
    			& $ 5.0_{-0.4}^{+0.4}$ & $ 1.99_{-0.03}^{+0.03}$ 	& $ 0.12_{-0.00}^{+0.01}$ 	& $ 8.7_{-0.2}^{+0.2}$ 	& $ 25_{-9}^{+14}$ 	& $ 1.12\,|\,889$ &  99.2\%\\  

  \\[-1.8ex]

\multicolumn{8}{c}{IC 342 X-6} \\
  \\[-1.8ex]
    0093640901	& $ 3.4_{-0.3}^{+0.3}$ & $ ... $ 					& $ 2.0_{-0.1}^{+0.2}$ 	& $ ... $ 				  	& $ 4.9_{-0.2}^{+0.2}$ 		& $ 1.14\,|\,179$ &  89.9\%\\ 
     			& $ 5.7_{-0.5}^{+0.6}$ & $ 1.65_{-0.08}^{+0.09}$ 	& $ ... $ 					& $ 5.9_{-0.2}^{+0.2}$ 	& $ ... $ 						& $ 0.88\,|\,179$ &  13.4\%\\ 

    2916		& $ 3.1_{-0.1}^{+0.5}$ & $ ... $ 					& $ 1.9_{-0.2}^{+0.2}$ 	& $ ... $ 				  	& $ 5.3_{-0.3}^{+0.3}$ 		& $ 1.16\,|\,80$ &  84.1\%\\ 
     			& $ 5.4_{-0.7}^{+0.7}$ & $ 1.63_{-0.12}^{+0.13}$ 	& $ ... $ 					& $ 6.5_{-0.2}^{+0.2}$ 	& $ ... $ 						& $ 0.79\,|\,80$ &   8.3\%\\ 

    2917		& $ 3.6_{-0.5}^{+0.5}$ & $ ... $ 					& $ 1.8_{-0.2}^{+0.2}$ 	& $ ... $ 				  	& $ 5.2_{-0.2}^{+0.3}$ 		& $ 1.51\,|\,86$ &  99.8\%\\ 
     			& $ 6.2_{-0.8}^{+0.8}$ & $ 1.70_{-0.13}^{+0.13}$ 	& $ ... $ 					& $ 6.7_{-0.2}^{+0.3}$ 	& $ ... $ 						& $ 1.21\,|\,86$ &  91.1\%\\ 

    0206890101	& $ 4.8_{-0.15}^{+0.2}$ & $ ... $ 					& $ 1.58_{-0.04}^{+0.04}$ 	& $ ... $ 				  	& $ 11.28_{-0.05}^{+0.06}$ 		& $ 1.22\,|\,798$ & 100.0\%\\ 
     			& $ 8.2_{-0.3}^{+0.3}$ & $ 1.99_{-0.03}^{+0.04}$ 	& $ ... $ 					& $ 16.0_{-0.4}^{+0.4}$ 	& $ ... $ 						& $ 0.97\,|\,798$ &  26.5\%\\ 
    			& $10.7_{-1.5}^{+1.4}$ & $ 2.10_{-0.07}^{+0.07}$ 	& $ 0.14_{-0.02}^{+0.02}$ 	& $ 18_{-1}^{+1}$ 	& $ 42_{-29}^{+62}$ 	& $ 0.96\,|\,796$ &  21.3\%\\ 

    0206890201	& $ 3.6_{-0.2}^{+0.2}$ & $ ... $ 					& $ 1.75_{-0.06}^{+0.06}$ 	& $ ... $ 				  	& $ 5.45_{-0.08}^{+0.08}$ 		& $ 1.47\,|\,524$ & 100.0\%\\ 
     			& $ 6.3_{-0.3}^{+0.3}$ & $ 1.81_{-0.05}^{+0.05}$ 	& $ ... $ 					& $ 7.03_{-0.15}^{+0.16}$ 	& $ ... $ 						& $ 1.01\,|\,524$ &  58.2\%\\ 
    			& $ 7.4_{-0.8}^{+1.1}$ & $ 1.70_{-0.10}^{+0.09}$ 	& $ 0.29_{-0.06}^{+0.10}$ 	& $ 6.7_{-0.4}^{+0.4}$ 	& $ 2.2_{-1.1}^{+2.9}$ 		& $ 0.98\,|\,522$ &  36.4\%\\ 

    0206890401	& $ 4.9_{-0.2}^{+0.3}$ & $ ... $ 					& $ 1.74_{-0.06}^{+0.07}$ 	& $ ... $ 				  	& $ 14.9_{-0.2}^{+0.2}$ 		& $ 1.23\,|\,429$ &  99.9\%\\
     			& $ 8.3_{-0.4}^{+0.4}$ & $ 1.87_{-0.05}^{+0.05}$ 	& $ ... $ 					& $ 19.8_{-0.6}^{+0.7}$ 	& $ ... $ 						& $ 1.07\,|\,429$ &  84.5\%\\ 
    			& $11.0_{-1.8}^{+1.8}$ & $ 2.00_{-0.09}^{+0.09}$ 	& $ 0.13_{-0.02}^{+0.02}$ 	& $ 22_{-2}^{+2}$ 	& $ 105_{-76}^{+198}$ 	& $ 1.05\,|\,427$ &  78.4\%\\ 
 
 \\[-1.8ex]

\multicolumn{8}{c}{Holmberg II X-1} \\
  \\[-1.8ex]
    0112520601	& $ 1.7_{-0.1}^{+0.1}$ & $ 2.71_{-0.04}^{+0.04}$ 	& $ ... $ 					& $ 25.0_{-0.7}^{+0.7}$ 	& $ ... $ 						& $ 1.03\,|\,652$ &  69.6\%\\ 
    			& $ 1.4_{-0.3}^{+0.4}$ & $ 2.5_{-0.2}^{+0.2}$ 	& $ 0.30_{-0.10}^{+0.06}$ 	& $ 11_{-2}^{+2}$ 	& $ 3.0_{-1.0}^{+1.0}$ 		& $ 1.00\,|\,650$ &  48.5\%\\ 

    0112520701	& $ 1.14_{-0.1}^{+0.1}$ & $ 2.42_{-0.05}^{+0.05}$ 	& $ ... $ 					& $ 19.2_{-0.6}^{+0.6}$ 	& $ ... $ 						& $ 1.03\,|\,503$ &  68.7\%\\ 
    			& $ 1.2_{-0.2}^{+0.3}$ & $ 2.2_{-0.1}^{+0.1}$ 	& $ 0.22_{-0.04}^{+0.05}$ 	& $ 9.4_{-0.8}^{+0.7}$ 	& $ 3.6_{-1.2}^{+2.1}$ 		& $ 0.97\,|\,501$ &  31.8\%\\ 

    0112520901	& $ 1.3_{-0.2}^{+0.2}$ & $ 3.0_{-0.1}^{+0.1}$ 	& $ ... $ 					& $ 5.4_{-0.4}^{+0.5}$  	& $ ... $ 						& $ 1.08\,|\,202$ &  79.9\%\\ 
    			& $ 1.5_{-0.4}^{+0.6}$ & $ 2.8_{-0.2}^{+0.2}$ 	& $ 0.16_{-0.04}^{+0.07}$ 	& $ 2.6_{-0.5}^{+0.5}$ 	& $ 1.9_{-1.3}^{+4.5}$ 		& $ 1.06\,|\,200$ &  71.8\%\\ 

    0200470101	& $ 1.70_{-0.03}^{+0.03}$ & $ 2.72_{-0.01}^{+0.01}$ 	& $ ... $ 					& $ 26.4_{-0.3}^{+0.3}$ 	& $ ... $ 						& $ 1.25\,|\,1338$ & 100.0\%\\ 
    			& $ 1.62_{-0.08}^{+0.08}$ & $ 2.55_{-0.04}^{+0.03}$ 	& $ 0.24_{-0.02}^{+0.02}$ 	& $ 12.7_{-0.5}^{+0.5}$ 	& $ 3.3_{-0.4}^{+0.5}$ 		& $ 1.12\,|\,1336$ &  99.9\%\\ 

  \\[-1.8ex]

\multicolumn{8}{c}{Holmberg IX X-1} \\
  \\[-1.8ex]
    0112521001	& $ 1.5_{-0.1}^{+0.1}$ & $ 1.79_{-0.03}^{+0.03}$ 	& $ ... $ 					& $ 11.5_{-0.1}^{+0.1}$ 	& $ ... $ 						& $ 1.06\,|\,813$ &  87.4\%\\ 
    			& $ 1.8_{-0.2}^{+0.3}$ & $ 1.67_{-0.06}^{+0.05}$ 	& $ 0.27_{-0.04}^{+0.06}$ 	& $ 10.9_{-0.4}^{+0.3}$ 	& $ 2.0_{-0.6}^{+1.0}$ 		& $ 1.02\,|\,811$ &  63.0\%\\ 

    0112521101	& $ 1.8_{-0.1}^{+0.1}$ & $ 1.85_{-0.03}^{+0.03}$ 	& $ ... $ 					& $ 13.8_{-0.2}^{+0.2}$ 	& $ ... $ 						& $ 0.97\,|\,908$ &  28.3\%\\ 
    			& $ 2.0_{-0.2}^{+0.3}$ & $ 1.77_{-0.06}^{+0.05}$ 	& $ 0.26_{-0.05}^{+0.07}$ 	& $ 13.3_{-0.5}^{+0.4}$ 	& $ 1.9_{-0.7}^{+1.3}$ 		& $ 0.95\,|\,906$ &  15.1\%\\ 

    3786		& $ 1.4_{-0.1}^{+0.1}$ & $ 1.73_{-0.04}^{+0.05}$ 	& $ ... $ 					& $ 11.76_{-0.05}^{+0.06}$ 	& $ ... $ 						& $ 1.16\,|\,249$ &  95.9\%\\ 
    			& $ 1.8_{-0.4}^{+0.5}$ & $ 1.61_{-0.12}^{+0.09}$ 	& $ 0.25_{-0.06}^{+0.10}$ 	& $ 11.5_{-0.5}^{+0.4}$ 	& $ 1.7_{-0.7}^{+1.5}$ 		& $ 1.09\,|\,247$ &  84.0\%\\ 

    3787		& $ 2.1_{-0.1}^{+0.1}$ & $ 1.92_{-0.04}^{+0.04}$ 	& $ ... $ 					& $ 16.6_{-0.2}^{+0.2}$ 	& $ ... $ 						& $ 1.12\,|\,259$ &  90.6\%\\ 

    3788		& $ 1.5_{-0.15}^{+0.15}$ & $ 1.70_{-0.05}^{+0.05}$ 	& $ ... $ 					& $ 10.27_{-0.06}^{+0.07}$ 	& $ ... $ 						& $ 1.14\,|\,215$ &  92.5\%\\ 
    			& $ 2.2_{-0.5}^{+0.7}$ & $ 1.55_{-0.13}^{+0.11}$ 	& $ 0.23_{-0.05}^{+0.08}$ 	& $ 10.0_{-0.5}^{+0.4}$ 	& $ 2.3_{-0.9}^{+2.1}$ 		& $ 1.02\,|\,213$ &  60.4\%\\ 

    0200980101	& $ 1.07_{-0.03}^{+0.03}$ & $ 1.61_{-0.01}^{+0.01}$ 	& $ ... $ 					& $ 9.30_{-0.02}^{+0.02}$ 	& $ ... $ 						& $ 1.46\,|\,2088$ & 100.0\%\\ 
    			& $ 1.8_{-0.1}^{+0.1}$ & $ 1.48_{-0.02}^{+0.02}$ 	& $ 0.24_{-0.01}^{+0.01}$ 	& $ 9.04_{-0.08}^{+0.08}$ 	& $ 2.7_{-0.3}^{+0.4}$ 		& $ 1.12\,|\,2086$ & 100.0\%\\

  \\[-1.8ex]

\multicolumn{8}{c}{NGC 5204 X-1} \\
  \\[-1.8ex]
    0142770101	& $ 0.5_{-0.1}^{+0.1}$ & $ 2.05_{-0.05}^{+0.05}$ 	& $ ... $ 					& $ 3.68_{-0.07}^{+0.07}$ 	& $ ... $ 						& $ 0.98\,|\,501$ &  36.5\%\\ 
   				& $ 0.7_{-0.2}^{+0.3}$ & $ 1.92_{-0.08}^{+0.08}$ 	& $ 0.23_{-0.04}^{+0.06}$ 	& $ 3.4_{-0.2}^{+0.2}$ 	& $ 0.9_{-0.3}^{+0.6}$ 		& $ 0.94\,|\,499$ &  16.2\%\\ 

    0142770301	& $ 1.0_{-0.2}^{+0.2}$ & $ 2.33_{-0.09}^{+0.09}$ 	& $ ... $ 					& $ 5.3_{-0.2}^{+0.3}$ 	& $ ... $ 						& $ 1.01\,|\,208$ &  56.0\%\\ 
    			& $ 1.7_{-0.6}^{+0.8}$ & $ 2.15_{-0.15}^{+0.14}$ 	& $ 0.17_{-0.03}^{+0.05}$ 	& $ 4.9_{-0.5}^{+0.5}$ 	& $ 4.9_{-2.7}^{+7.5}$ 		& $ 0.90\,|\,206$ &  14.9\%\\ 

    0150650301	& $ 1.2_{-0.2}^{+0.2}$ & $ 2.48_{-0.08}^{+0.08}$ 	& $ ... $ 					& $ 6.3_{-0.3}^{+0.3}$ 	& $ ... $ 						& $ 1.01\,|\,295$ &  56.7\%\\ 
    			& $ 1.2_{-0.6}^{+0.6}$ & $ 2.2_{-0.3}^{+0.2}$ 	& $ 0.25_{-0.08}^{+0.11}$ 	& $ 5.2_{-1.2}^{+0.8}$ 	& $ 1.9_{-0.8}^{+2.0}$ 		& $ 0.97\,|\,293$ &  35.5\%\\ 

    3933		& $ 1.6_{-0.15}^{+0.15}$ & $ 2.94_{-0.08}^{+0.08}$ 	& $ ... $ 					& $ 3.8_{-0.2}^{+0.2}$ 	& $ ... $ 						& $ 1.64\,|\,147$ & 100.0\%\\ 
    			& $ 1.7_{-0.3}^{+0.3}$ & $ 2.67_{-0.14}^{+0.13}$ 	& $ 0.19_{-0.03}^{+0.04}$ 	& $ 3.0_{-0.4}^{+0.4}$ 	& $ 1.4_{-0.5}^{+0.8}$ 		& $ 1.45\,|\,145$ & 100.0\%\\ 

    0405690101	& $ 1.6_{-0.1}^{+0.1}$ & $ 2.73_{-0.05}^{+0.05}$ 	& $ ... $ 					& $ 9.0_{-0.3}^{+0.3}$ 	& $ ... $ 						& $ 1.07\,|\,584$ &  89.6\%\\ 
    			& $ 1.5_{-0.2}^{+0.2}$ & $ 2.6_{-0.1}^{+0.1}$ 	& $ 0.23_{-0.06}^{+0.07}$ 	& $ 8.0_{-0.9}^{+0.7}$ 	& $ 1.7_{-0.8}^{+1.4}$ 		& $ 1.06\,|\,582$ &  85.1\%\\ 

    0405690201	& $ 1.52_{-0.07}^{+0.07}$ & $ 2.62_{-0.03}^{+0.03}$ 	& $ ... $ 					& $ 7.5_{-0.2}^{+0.2}$ 	& $ ... $ 						& $ 1.15\,|\,818$ &  99.8\%\\ 
    			& $ 1.25_{-0.15}^{+0.15}$ & $ 2.34_{-0.09}^{+0.08}$ 	& $ 0.29_{-0.03}^{+0.03}$ 	& $ 5.8_{-0.5}^{+0.5}$ 	& $ 1.8_{-0.4}^{+0.3}$ 		& $ 1.08\,|\,816$ &  94.2\%\\ 

    0405690501	& $ 0.84_{-0.08}^{+0.08}$ & $ 2.27_{-0.04}^{+0.04}$ 	& $ ... $ 					& $ 4.7_{-0.1}^{+0.1}$ 	& $ ... $ 						& $ 1.09\,|\,669$ &  94.2\%\\ 
    			& $ 0.7_{-0.2}^{+0.2}$ & $ 1.9_{-0.1}^{+0.1}$ 	& $ 0.28_{-0.03}^{+0.03}$ 	& $ 3.7_{-0.3}^{+0.2}$ 	& $ 1.7_{-0.3}^{+0.3}$ 		& $ 0.92\,|\,667$ &   7.5\%\\

  \\[-1.8ex]

\multicolumn{8}{c}{NGC 5408 X-1} \\
  \\[-1.8ex]
    0112290501	& $ 1.3_{-0.2}^{+0.2}$ & $ 3.5_{-0.1}^{+0.1}$ 	& $ ... $ 					& $ 10.1_{-1.0}^{+1.1}$ 	& $ ... $ 						& $ 1.28\,|\,270$ &  99.9\%\\ 
   				& $ 1.0_{-0.4}^{+0.4}$ & $ 2.7_{-0.2}^{+0.2}$ 	& $ 0.18_{-0.02}^{+0.03}$ 	& $ 4.9_{-1.0}^{+1.1}$ 	& $ 8.8_{-2.5}^{+5.0}$ 		& $ 1.06\,|\,268$ &  75.9\%\\ 

    0112290601	& $ 1.4_{-0.2}^{+0.2}$ & $ 3.5_{-0.1}^{+0.1}$ 	& $ ... $ 					& $ 10.1_{-1.0}^{+1.1}$ 	& $ ... $ 						& $ 1.09\,|\,278$ &  85.6\%\\ 
    			& $ 0.7_{-0.3}^{+0.4}$ & $ 2.5_{-0.3}^{+0.3}$ 	& $ 0.22_{-0.03}^{+0.03}$ 	& $ 3.9_{-0.9}^{+1.1}$ 	& $ 6.7_{-1.4}^{+2.1}$ 		& $ 0.89\,|\,276$ &   8.5\%\\ 

    0112290701	& $ 0.9_{-0.2}^{+0.3}$ & $ 3.21_{-0.15}^{+0.16}$ 	& $ ... $ 					& $ 8.7_{-0.9}^{+1.1}$ 	& $ ... $ 						& $ 1.13\,|\,141$ &  86.9\%\\ 
    			& $ 0.8_{-0.5}^{+0.6}$ & $ 2.7_{-0.3}^{+0.3}$ 	& $ 0.18_{-0.04}^{+0.05}$ 	& $ 5.5_{-1.4}^{+1.5}$ 	& $ 6.8_{-2.9}^{+7.7}$ 		& $ 0.99\,|\,139$ &  49.2\%\\ 

    0112291201	& $ 1.2_{-0.3}^{+0.3}$ & $ 3.00_{-0.16}^{+0.17}$ 	& $ ... $ 					& $ 6.1_{-0.6}^{+0.8}$ 	& $ ... $ 						& $ 1.35\,|\,143$ &  99.6\%\\ 
    			& $ 1.2_{-0.6}^{+0.7}$ & $ 2.2_{-0.3}^{+0.3}$ 	& $ 0.19_{-0.04}^{+0.05}$ 	& $ 3.4_{-0.7}^{+0.8}$ 	& $ 6.2_{-2.2}^{+5.6}$ 		& $ 1.05\,|\,141$ &  66.2\%\\ 

    0302900101	& $ 1.35_{-0.1}^{+0.1}$ & $ 2.66_{-0.04}^{+0.04}$ 	& $ 0.17_{-0.01}^{+0.01}$ 	& $ 4.8_{-0.2}^{+0.2}$ 	& $ 8.4_{-0.9}^{+1.1}$ 		& $ 1.30\,|\,978$ & 100.0\%\\ 

  \\[-1.8ex]

\multicolumn{8}{c}{NGC 6946 X-6} \\
  \\[-1.8ex]
    1043		& $ 2.2_{-0.2}^{+0.2}$ & $ 2.5_{-0.1}^{+0.1}$ 	& $ ... $ 					& $ 4.7_{-0.2}^{+0.2}$ 	& $ ... $ 						& $ 1.57\,|\,165$ & 100.0\%\\ 
    			& $ 3.7_{-0.6}^{+0.6}$ & $ 2.2_{-0.1}^{+0.1}$ 	& $ 0.16_{-0.02}^{+0.02}$ 	& $ 4.1_{-0.3}^{+0.4}$ 	& $ 7.1_{-2.8}^{+4.9}$ 		& $ 0.95\,|\,163$ &  33.4\%\\ 

    4404		& $ 2.3_{-0.2}^{+0.3}$ & $ 2.4_{-0.1}^{+0.1}$ 	& $ ... $ 					& $ 4.3_{-0.2}^{+0.2}$ 	& $ ... $ 						& $ 1.59\,|\,117$ & 100.0\%\\ 
    			& $ 4.1_{-0.9}^{+1.0}$ & $ 2.2_{-0.2}^{+0.2}$ 	& $ 0.16_{-0.02}^{+0.03}$ 	& $ 3.8_{-0.4}^{+0.5}$ 	& $ 7.7_{-3.9}^{+9.2}$ 		& $ 1.12\,|\,115$ &  82.6\%\\ 

    0200670301	& $ 2.6_{-0.3}^{+0.3}$ & $ 2.8_{-0.1}^{+0.1}$ 	& $ ... $ 					& $ 5.2_{-0.4}^{+0.5}$ 	& $ ... $ 						& $ 1.19\,|\,166$ &  95.2\%\\ 
    			& $ 3.1_{-0.7}^{+0.9}$ & $ 2.4_{-0.2}^{+0.2}$ 	& $ 0.20_{-0.04}^{+0.05}$ 	& $ 3.7_{-0.7}^{+0.8}$ 	& $ 5.8_{-2.6}^{+7.5}$ 		& $ 0.98\,|\,164$ &  43.7\%\\ 

    4631		& $ 2.4_{-0.3}^{+0.3}$ & $ 2.7_{-0.1}^{+0.2}$ 	& $ ... $ 					& $ 4.0_{-0.3}^{+0.4}$ 	& $ ... $ 						& $ 1.58\,|\,100$ & 100.0\%\\ 
    			& $ 4.2_{-1.0}^{+1.2}$ & $ 2.3_{-0.2}^{+0.2}$ 	& $ 0.16_{-0.03}^{+0.04}$ 	& $ 3.3_{-0.5}^{+0.6}$ 	& $ 8.3_{-4.5}^{ +12}$ 	& $ 1.11\,|\,98$ &  79.2\%\\ 

    4632		& $ 2.2_{-0.3}^{+0.3}$ & $ 2.5_{-0.1}^{+0.1}$ 	& $ ... $ 					& $ 4.4_{-0.3}^{+0.3}$ 	& $ ... $ 						& $ 1.94\,|\,102$ & 100.0\%\\ 
    			& $ 4.9_{-1.0}^{+1.1}$ & $ 2.3_{-0.2}^{+0.2}$ 	& $ 0.14_{-0.02}^{+0.03}$ 	& $ 4.0_{-0.5}^{+0.6}$ 	& $ 17_{-9}^{ +23}$ 	& $ 1.36\,|\,100$ &  99.0\%\\ 

    4633		& $ 2.2_{-0.3}^{+0.3}$ & $ 2.6_{-0.1}^{+0.1}$ 	& $ ... $ 					& $ 4.3_{-0.3}^{+0.3}$ 	& $ ... $ 						& $ 1.71\,|\,107$ & 100.0\%\\ 
    			& $ 4.6_{-1.0}^{+1.1}$ & $ 2.2_{-0.2}^{+0.2}$ 	& $ 0.14_{-0.02}^{+0.03}$ 	& $ 3.8_{-0.5}^{+0.6}$ 	& $ 14_{-7}^{ +19}$ 	& $ 1.05\,|\,105$ &  66.0\%\\ 

    0401360301	& $ 2.5_{-0.5}^{+0.5}$ & $ 2.7_{-0.2}^{+0.2}$ 	& $ ... $ 					& $ 5.8_{-0.8}^{+1.1}$ 	& $ ... $ 						& $ 1.28\,|\,62$ &  93.2\%\\ 
    			& $ 3.0_{-1.2}^{+1.5}$ & $ 2.3_{-0.6}^{+0.4}$ 	& $ 0.21_{-0.07}^{+0.11}$ 	& $ 4.2_{-1.4}^{+1.6}$ 	& $ 5.6_{-3.7}^{ +22}$ 	& $ 1.19\,|\,60$ &  84.7\%\\

  \\[-1.8ex]

\footnotetext[1]{Hydrogen column density in units $10^{21}$ cm$^{-2}$.}
\footnotetext[2]{Photon index.}
\footnotetext[3]{Inner disk temperature of the {\sc diskbb} component in units of keV.}
\footnotetext[4]{Power-law model luminosity in 0.4--10 keV band in units of $10^{39}\,\ergs$.}
\footnotetext[5]{{\sc diskbb} model bolometric luminosity (computed in 0.01--50 keV band) in units of $10^{39}\,\ergs$.}
\footnotetext[6]{Reduced $\chi^2$ and the degrees of freedom.}
\footnotetext[7]{Rejection probability.}

\end{longtable}


\label{lastpage}


\end{document}